# Orbit-spin coupling and the interannual variability of global-scale dust storm occurrence on Mars

James H. Shirley and Michael A. Mischna

*Jet Propulsion Laboratory, Pasadena, California 91109 USA*

4 May 2016

*Key points*:

A weak coupling of the orbital and rotational angular momenta of Mars may significantly perturb the circulation of the atmosphere

We test this hypothesis through a comparison of dynamical forcing functions for Mars years with and without global-scale dust storms

All known global-scale dust storms on Mars occurred in years in which the hypothesis 'retrodicts' an intensification of circulatory flows

**Headings:**

**Abstract:**

A new physical hypothesis predicts that a weak coupling of the orbital and rotational motions of extended bodies may give rise to a modulation of circulatory flows within their atmospheres. Driven cycles of intensification and relaxation of large-scale circulatory flows are predicted, with the phasing of these changes linked directly to the rate of change of the orbital angular momentum, *dL/dt*, with respect to inertial frames. We test the hypothesis that global-scale dust storms (GDS) on Mars may occur when periods of circulatory intensification (associated with positive and negative extrema of the *dL/dt* waveform) coincide with the southern summer dust storm season on Mars. The orbit-spin coupling hypothesis additionally predicts that the intervening 'transitional' periods, which are characterized by the disappearance and subsequent sign change of *dL/dt*, may be unfavorable for the occurrence of GDS, when they occur during the southern summer dust storm season. These hypotheses are confirmed through comparisons between calculated dynamical time series of *dL/dt* and historic observations. All of the nine known global-scale dust storms on Mars took place during Mars years when circulatory intensification during the dust storm season is predicted under the orbit-spin coupling hypothesis. No historic global-scale dust storms were recorded during transitional intervals. Orbit-spin coupling accelerations evidently contribute to the interannual variability of the Mars atmosphere.

# 1. Introduction

Small-scale to regional-scale dust storms are observed on Mars during all seasons of the year [*Martin and Zurek*, 1993; *Cantor et al.*, 2001; *Wang and Richardson*, 2015]. A fairly complicated annual dust cycle is likewise seen; multiple episodes of increased atmospheric dust loading typically occur during the southern summer season on Mars, when the planet is near perihelion and when the solar input to the atmospheric system is at a maximum [*Kass et al.*, 2014; *Wang and Richardson*, 2015; *Montabone et al.*, 2015; *Lemmon et al.*, 2015]. The suspended dust absorbs, scatters, and re-radiates the solar radiation, thereby locally heating the atmosphere and modifying convective activity and circulatory flows.

At irregular intervals of some years, and for unknown reasons, multiple small and regional-scale dust storms may explosively grow and coalesce, leading to the obscuration of surface features of Mars by bright dust clouds of hemispheric and even global scale. When the Mariner 9 spacecraft arrived at Mars in 1971, the planet's surface was found to be almost entirely obscured by a dust storm that persisted for several months before disappearing [*Martin*, 1974]. Global-scale dust storm (GDS) events also occurred one Mars year later in 1973, and then again in 1977 [*Zurek et al.*, 1992]. The two most recent such storms occurred in 2001 and in 2007; these were recorded by the Mars Global Surveyor spacecraft [*Albee et al.*, 1998] and the Mars Reconnaissance Orbiter spacecraft [*Zurek and Smrekar*, 2007], respectively. Important aspects of these storms are described in *Smith et al.* [2002], *Strausberg et al.* [2005], *Cantor* [2007], *Cantor et al.* [2008], *Kass et al.* [2007], *Kleinböhl et al.* [2009], *Martinez-Alvarado et al.* [2009], *Clancy et al.* [2010], *Guzewich et al.* [2014], and *Wang and Richardson* [2015].

Global-scale dust storms on Mars represent the most spectacular form of atmospheric interannual variability observed on any of the terrestrial planets. During these events, atmospheric temperatures may locally exceed their seasonal norms by more than 40 K (although at the surface, a moderation of diurnal temperature extremes is typically observed, due to the interception of incoming solar radiation and outgoing thermal radiation by atmospheric dust). The atmosphere, thus warmed, expands significantly. Dust may extend to altitudes of more than 60 km above the surface, and the large-scale circulation is profoundly impacted [*Gierasch and Goody*, 1973; *Leovy et al.*, 1973; *Zurek*, 1981a, b; *Haberle et al.*, 1982; *Haberle*, 1986; *Kahn et al.*, 1992; *Leovy*, 2001; *Clancy et al.*, 1994, 2000; *Heavens et al.*, 2011a, 2011b; *Medvedev et al.*, 2011; *Guzewich et al.*, 2013]. It appears that both surface wind stresses and the large-scale circulation of the atmosphere must intensify in order to produce the explosive growth of global-scale dust storms [*Haberle*, 1986; *Shirley*, 2015, and references therein], but the source or sources of such an intensification have remained elusive. While a large number of possible contributing physical processes and interactions have been proposed [*cf. Leovy et al.*, 1973; *Haberle*, 1986; *Zurek et al.*, 1992; *Pankine and Ingersoll*, 2002, 2004; *Fenton et al.*, 2006; *Rafkin*, 2009], prior attempts to account for the interannual variability of GDS events through numerical modeling have not met with great success [*cf. Pankine and Ingersoll*, 2002; *Basu et al.*, 2006; *Rafkin*, 2009; *Mulholland et al.*, 2013; *Newman and Richardson*, 2015].

Presently the leading explanation for the occurrence of GDS in some Mars years and not in others invokes an atmospheric memory component that is supplied by a variable spatial distribution (and hence supply) of dust. Surface dust redistribution, resulting in a depletion of surface dust reservoirs in key high-wind-stress locations in a given year, may act to inhibit the formation and growth of GDS in subsequent years [*Haberle*, 1986; *Basu et al.*, 2006; *Pankine and Ingersoll*, 2004; *Fenton et al.*, 2006; *Cantor et al.*, 2007; *Mullholland et al.*, 2013; *Newman* *and Richardson*, 2015]. Fuller discussion of these topics is found in the cited references and in *Shirley* [2015].

Recently, it has been suggested that solar system dynamics may play a role in the initiation of global-scale dust storms on Mars. *Shirley* [2015] (hereinafter [S15]) found systematic relationships linking the occurrence and non-occurrence of GDS with changes in the orbital angular momentum of Mars with respect to the barycenter of the solar system. All of the global-scale dust storms in the historic record grew to planet-encircling status during periods when the orbital angular momentum of Mars, $L_{Mars}$, was increasing or near peak values. An opposite tendency was noted for global-storm-free years. In addition, systematic phase leads and lags (of the $L_{Mars}$ waveform with respect to solar irradiance) were found for storms that occurred anomalously early and unusually late in the dust storm season. Physical mechanisms were not discussed in [S15].

In a subsequent investigation, *Shirley* [2016] (hereinafter [S16]) obtained a mathematical expression describing a weak coupling between the orbital and rotational motions of extended bodies. Accelerations resulting from this coupling are predicted to give rise to cycles of intensification and relaxation of circulatory flows within planetary atmospheres. In the present paper, we make use of the catalog of global-scale dust storms presented in [S15] in order to test theoretical predictions made in [S16].

In Section 2 we describe the updated catalog of GDS years and global-storm-free Mars years that is employed for hypothesis testing. Section 3 briefly describes the physical mechanism introduced in [S16] and identifies the linked hypotheses that are tested in the present investigation. We describe the calculation of the dynamical forcing function $dL/dt$ for Mars for the interval from 1920-2030, together with the method employed to quantify the phasing of this waveform with respect to the annual cycle of solar irradiance. A method for evaluating the statistical significance of the resulting distributions of angular phases is also described in Section 3. Results are reported in Section 4 and are discussed and interpreted in Section 5. Section 6 lists caveats and discusses some of the implications. Forecasts for the next two dust storm seasons on Mars are presented in Section 7, while Section 8 summarizes the conclusions of this investigation.

**2. Observations: Mars years with (and without) global-scale dust storms**

Mars has been under continuous observation by orbiting spacecraft since 1997. The subsequent period (through 2015) includes 10 Mars years (Table 1). Global-scale dust storms occurred in two Mars years (Mars Year 25 and Mars Year 28) during this period (the numbering convention for Mars years was introduced in *Clancy et al.*, [2000], and was recently updated for years prior to MY 1 (1955) by *Piqueux et al.* [2015]). Spacecraft observations also document a number of GDS occurring in the decade of the 1970s. Due to the difficulty of observing the entire period of Mars' dust storm season from Earth [*Zurek and Martin*, 1993; S15]), the number of years prior to 1970 that may be categorized as GDS years or GDS-free years is quite limited. Further, the inception dates of a number of the global storms are uncertain [S15]. Methods and criteria employed in the compilation of Table 1 are described in more detail in [S15]. For the present investigation we have added the recently completed Mars year 32 to the [S15] catalog.

| Perihelion Date | Mars Year | Description | GDS start date ($L_s$) |
|---|---|---|---|
| 1924.654 | -16 | Solstice season global dust storm | 310.0 |
| 1939.714 | -8 | Global-scale storm free | |
| 1956.633 | 1 | Solstice season global dust storm | 249.0 |
| 1971.692 | 9 | Solstice season global dust storm | 260.0 |
| 1973.580 | 10 | Solstice season global dust storm | 300.0 |
| 1975.443 | 11 | Global-scale storm free | |
| 1977.330 | 12 | Equinox season global dust storm | 204.0 |
| 1982.971 | 15 | Solstice season global dust storm | 208.0 |
| 1986.722 | 17 | Global-scale storm free | |
| 1988.611 | 18 | Global-scale storm free | |
| 1994.251 | 21 | Solstice season global dust storm | 254.0 |
| 1998.029 | 23 | Global-scale storm free | |
| 1999.892 | 24 | Global-scale storm free | |
| 2001.782 | 25 | Equinox season global dust storm | 185.0 |
| 2003.673 | 26 | Global-scale storm free | |
| 2005.533 | 27 | Global-scale storm free | |
| 2007.423 | 28 | Solstice season global dust storm | 262.0 |
| 2009.311 | 29 | Global-scale storm free | |
| 2011.201 | 30 | Global-scale storm free | |
| 2013.062 | 31 | Global-scale storm free | |
| 2014.952 | 32 | Global-scale storm free | |

**Table 1**. Catalog of Mars years with and without global-scale dust storms employed for this study. Mars year identifications for years prior to 1955 follow the method of *Piqueux et al.* [2015]. "Solstice season" refers to Mars' southern summer solstice only. The 10-day time step employed for calculations limits the precision of the perihelion dates.

Two categories of global-scale dust storm years are identified in the third column of Table 1. Following [S15], seven of the nine GDS years are identified as (southern summer) solstice season storm years, while two others (MY 12 and MY 25) are labeled equinox season storm years. Attempts to reproduce equinox season storms within global circulation models have been largely unsuccessful [*cf. Basu et al.*, 2006; *Newman and Richardson*, 2015], presumably due to the relatively low levels of solar irradiance available to drive atmospheric motions at this season. In contrast, the southern summer solstice season storms more closely coincide in time with the period of maximum solar input to the Mars atmosphere, due to the near-coincidence of perihelion (at $L_s \sim 250°$) with the southern summer solstice (at $L_s=271°$; *cf.* Fig. 2 of S15). ($L_s$, the celestial longitude of the Sun as viewed from Mars, describes the progression of the seasons on Mars in a manner analogous to the seasons of Earth. In both cases, the vernal equinox is identified with the beginning of the northern spring season. $L_s$ is assigned a zero value at the time of the vernal equinox).

**3. Orbit-spin coupling: A possible forcing function for atmospheric variability on seasonal and longer time scales**

A new approach to the problem of a coupling of the orbital and spin motions of extended bodies is developed in *Shirley* [2016]. Through the introduction of the phenomenon of barycentric revolution within the context of a classical derivation of inertial forces, [S16] arrives at a previously unknown coupling expression that does not involve gravitational or thermal tides. The coupling locally takes the form of an acceleration, hereafter labeled the "*coupling term acceleration*," or CTA, which is given by

$$\text{CTA} = -c\,(\dot{\boldsymbol{L}} \times \boldsymbol{\omega}_\alpha) \times \boldsymbol{r}, \qquad (1)$$

where $\dot{\boldsymbol{L}}$ is the time rate of change of the orbital angular momentum ($d\boldsymbol{L}/dt$) with respect to inertial frames, $\boldsymbol{\omega}_\alpha$ is the angular velocity of rotation, and $\boldsymbol{r}$ is a position vector originating at the center of the subject body. The leading coefficient, $c$, is a coupling efficiency factor whose value will depend on the physical makeup of the subject body and the physical system under investigation. Equation 1 describes an acceleration field that varies with position and with time on and within a subject body (*cf.* Fig. 5 of [S16]).

For the present investigation we presume that $c \neq 0$ and that the accelerations are of a small but appreciable magnitude in comparison with other forces acting. (A preliminary numerical value for $c$ in the case of the Mars atmosphere is obtained in the companion paper by *Mischna and Shirley* [2016]). We recognize here that the variability of the coupling term acceleration (CTA) on seasonal time scales is dictated largely if not entirely by *d**L**/dt*. *d**L**/dt* may thus be considered to represent a *forcing function*, whose variability with time may usefully be compared with observations. Our method for calculation of *d**L**/dt* is described in Section 3.2 below.

Potentially observable consequences of the acceleration field of Equation 1 are detailed in [S16]. We next summarize two key predictive statements made in [S16] that inform the present investigation.

3.1. Predictions and hypothesis testing

3.1.1. Constructive and destructive interference with circulatory motions

We first presume the existence of a planetary atmosphere, with a pre-existing global wind field, whose circulatory motions are determined by atmospheric and surface responses to solar heating, by large-scale topography, and by surface properties, among other known factors. The coupling term accelerations (CTA) of Equation 1 will modify the pre-existing velocities of the particles or parcels of atmosphere at all locations and altitudes where CTA$\neq$0. The accelerations will thereby constructively or destructively interfere with the atmospheric motions that would otherwise take place in the absence of the CTA. This is the most general predictive statement that could be advanced in [S16]. However, this consequence is not well suited to evaluation and testing simply by means of comparison with a set of observations with annual resolution (as in Table 1).

3.1.2. Cycles of intensification and relaxation of circulatory flows

In a comparison between otherwise identical atmospheres, one with and one without the accelerations, it seems inevitable that the accelerated atmosphere must exhibit driven cycles of overall intensification and relaxation of circulatory flows. This implies that (to first order) the peaks of the intensification periods for the subject atmosphere should correspond in time to the extrema of the *dL/dt* waveform, while the relaxation periods will correspond to the intervals bracketing the times when *dL/dt* approaches and attains zero values. (In Section 3.2 we will calculate and display the *dL/dt* waveform and more fully describe these effects).

We must here caution against a too-literal interpretation of the term 'intensification.' As discussed in some detail in [S16], the accelerations are not uniform everywhere, but instead exhibit substantial variability as a function of latitude, longitude, and time. While it is possible that a linear or monotonic "speeding up" of some pre-existing circulatory flows might occasionally occur, it is far more likely that the adjustment of the atmosphere will take the form of modified patterns of surface winds and atmospheric pressures, leading to morphological changes on larger scales. Simple linear relationships of atmospheric observables to altered forcing are likely to be the exception, rather than the rule.

3.1.3. Hypotheses for testing

The occurrence of global-scale dust storms in the Mars atmosphere appears to require intensification of both surface wind stresses and of large-scale circulatory flows [*Haberle*, 1986;

see also S15, and references therein]. The newly proposed physical mechanism of Equation 1 above appears well suited to provide the requisite modulation of circulatory flows, if the accelerations attain magnitudes sufficient to perturb the large-scale circulation. We thus arrive at the following paired hypotheses, which may be tested through a comparison of the phasing of the calculated dynamical forcing function with historic observations of past global-scale dust storms:

**H1a**. Coincidence in time of the Mars dust storm season with intervals bracketing the zero-crossings of the *dL/dt* waveform (i.e., the 'circulatory relaxation periods') should be *unfavorable* for the occurrence of global-scale dust storms. Conversely:

**H1b**. Global-scale dust storms are more likely to occur in Mars years when intervals of intensification of the global circulation coincide with the southern summer dust storm season.

The 'exclusionary' hypothesis (**H1a**) provides the more stringent test, as the occurrence of some number of GDS events within predicted circulatory relaxation intervals would argue against the relevance of the mechanism suggested. With reference to **H1b**, we recognize that not all years experiencing circulatory intensification during the dust storm season need necessarily include global-scale dust storms. There are a number of scenarios that might preclude the development of a global-scale storm in any given year. Among these are: 1) an unfavorable pre-existing spatial distribution of surface dust, or 2) insufficient amplitudes of the coupling term accelerations to trigger GDS activity.

Evaluation of hypotheses **H1a** and **H1b** is the primary objective of the present investigation.

3.2. Calculation of *dL/dt* and its relationship to the annual cycle of solar irradiance

An extended discussion of the solar system dynamical processes giving rise to the variability with time of the barycentric orbital angular momentum of Mars is provided in Section 3 of [S15]. That introduction will not be repeated here; instead, in this sub-section, we provide an outline of the method employed to obtain the forcing function *dL/dt*. As in [S15], we first obtain the instantaneous orbital angular momentum of Mars with respect to the solar system barycenter, using the following equation, from *Jose* [1965];

$$\mathbf{L} = [(y\dot{z} - z\dot{y})^2 + (z\dot{x} - x\dot{z})^2 + (x\dot{y} - y\dot{x})^2]^{1/2} \tag{2}$$

Here, the required quantities are the positional coordinates (*x, y, z*) and velocities ($\dot{x}, \dot{y}, \dot{z}$) of the subject body with respect to the solar system barycenter. The mass is not explicitly included (but must be supplied later as a multiplicative factor for quantitative comparisons). The positions and velocities required may be obtained from JPL's online Horizons ephemeris system [*Giorgini et al.*, 1996; *Giorgini*, 2015]. To obtain the time derivative of (2) it is simplest to merely difference the values of each of the vector components of **L** for two adjacent times, and divide each resulting component by the time difference. We then assign an intermediate time value to the 3-component rate of change vector obtained. The series of *dL/dt* at 10-day intervals for the study period 1920-2030 is available in the online Supplementary Data.

The orbital angular momentum vector lies in a direction normal to the instantaneous orbit plane. While its time derivative (*dL/dt*) is not similarly constrained, in practice we find that the *z*-component of this quantity (in ecliptic coordinates) is almost always considerably larger than the accompanying *x*- and *y*-components (that lie more nearly within the orbit plane). Thus for plotting purposes it is sufficient to employ only the signed ecliptic *z*-component to adequately represent the magnitude of the variability with time of *dL/dt*. Figure 1 displays the signed *z*-component, in ecliptic coordinates, of the rate of change of the barycentric orbital angular momentum (*dL/dt*) of Mars for an interval from late 2006 to early 2017.

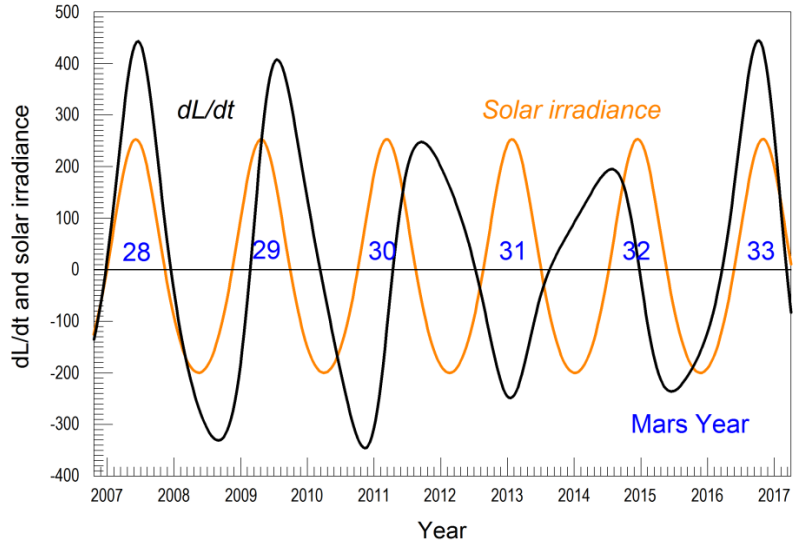

**Figure 1**. Phasing of the rate of change of the orbital angular momentum *dL/dt* (in black) with respect to the annual cycle of solar irradiance (orange curve) over 5.5 Mars years (MY 28-33, from late 2006 to early 2017). Both curves have been arbitrarily scaled in order to better illustrate the relationships. Peaks of the irradiance occur during the southern summer season, at the time of Mars' perihelion, at $L_s$~250°. The mean value of the solar irradiance at Mars (~590 W m$^{-2}$; [S15]) corresponds to the zero value on the y-axis of this figure. This value is attained about 10 sols prior to the nominal start time of the dust storm season on Mars, at $L_s$ ~160° [Zurek and Martin, 1993; S15]. A global-scale dust storm occurred during MY 28; however, the subsequent years (through MY 32) have been free of such events. Solar system dynamics gives rise to the phase and amplitude modulation of *dL/dt* [S15]. The solar irradiance and *dL/dt* series employed are available in the online Supplementary Data.

Figure 1 also illustrates the variability with time of the solar irradiance received by Mars over the same time interval. Changes in the distance between Mars and the Sun due to the large eccentricity (*e*=0.09) of Mars' orbit account for the regular annual cycle displayed in orange in Fig. 1. As noted earlier, the irradiance peaks at perihelion, during late southern spring on Mars, when the orbital longitude of the Sun $L_s$ reaches ~250°. The dust storm season on Mars [*Zurek and Martin*, 1993; S15] is approximately symmetric around the time of perihelion. The dust storm season on Mars thus corresponds fairly closely to the portions of the orange curve lying above the zero line of Fig. 1. The time span of Fig. 1 has been chosen to include the perihelia (and hence the dust storm seasons) of six Mars years, from MY 28 through MY 33.

3.2.1. Classification of phase relationships in terms of waveform polarity

Mars gains orbital angular momentum, at the expense of other members of the solar system family, during intervals when *dL/dt* is positive [S15; S16]. Conversely, during intervals when *dL/dt* falls below the zero line of Fig. 1, Mars is instead yielding up orbital angular momentum to other members of the solar system. This is an important distinction that will play a role in the discussions of Section 5 below. Two key examples are illustrated in Fig. 1. For identification purposes, cases such as that of Mars year 28, shown at the left side of the figure, will be termed "positive polarity" years. Here a local maximum of the *dL/dt* waveform approximately coincides with the peak of the solar irradiance. Mars year 31, on the other hand, exhibits a diametrically opposed relationship of the two curves, with a local minimum of the *dL/dt* waveform coinciding with peak solar irradiance. Again, for convenience, cases similar to MY 31 will hereinafter be labeled "negative polarity" years.

Cycles of intensification and relaxation of circulatory flows of atmospheres are predicted to occur in phase with the variability of the *dL/dt* waveform as illustrated in Fig. 1 (Section 3.1). In particular, the positive and negative extrema of the curve are identified with predicted intervals of intensification of the large-scale circulation, while the intervening periods, when *dL/dt* approaches and transitions through zero values, are suspected to represent times of relaxation of circulatory flows. During these latter intervals, the putative driving accelerations disappear, to subsequently re-emerge with opposed directions [S16]. These episodes are labeled "transition intervals." If such an interval is centered near the midpoint of the dust storm season, we may describe this as a "transition year." Mars year 32 as illustrated in Fig. 1 is a good example; in this case the *dL/dt* waveform attains a zero amplitude very close to the midpoint of the dust storm season, i.e., at Mars' perihelion. Under the physical hypothesis outlined in Section 3.1, we consider this condition to be unfavorable for the occurrence of global-scale dust storms.

The remaining Mars years of Fig. 1 (MY 29, 30, and 33) exhibit phase relationships that fall somewhere between the above highlighted cases (positive polarity in MY 28, negative polarity in MY 31, and transitional, or 'zero-crossing' phasing in MY 32). In order to move forward with our comparisons, we require an objective measure of the phasing of the waveforms of Fig. 1.

3.2.2. *dL/dt* phase determination

We define a phase for all points on the *dL/dt* waveform that is analogous to the sine function. In this scheme, the upward-crossing and downward-crossing zero-amplitude transitions are assigned phases of 0° and 180°, respectively, while the times when positive and negative extrema of the *dL/dt* waveform are attained are assigned phase values of 90° and 270°, respectively. Phases for times intermediate between these cases are linearly interpolated on the basis of the time difference with respect to the bracketing reference points. This scheme provides a reasonably uniform distribution of phase values over the 110-yr interval of our calculations (from 1920-2030). As in Fig. 1, we employ the signed *z*-component of *dL/dt* in ecliptic coordinates for defining the reference points. A listing of the phase values obtained (at intervals of 10 days) is available in the Supplementary Data.

For comparison purposes we characterize each Mars year on the basis of the *dL/dt* phase value *at the time of perihelion*. The phases thus determined are listed in the first column of Table 2, where the Mars years of our sample are now ordered in terms of increasing phase values. In the following the phase parameter will occasionally be denoted by $\varphi_{dL/dt}$. The last column of Table 2 provides the (positive or negative) peak amplitude of the *dL/dt* waveform for positive polarity years (such as MY 28 of Fig. 1) and for negative polarity years (such as MY 31 of Fig. 1). Amplitude values are not displayed for transition years, which are times when the *dL/dt* waveform passes through zero and changes polarity during the southern spring and summer dust storm season.

| ϕ dL/dt at perihelion | Mars Year | Description | GDS start date ($L_s$) | dL/dt peak amplitude* |
|---|---|---|---|---|
| 1.5 | 18 | Global-scale storm free | | |
| 1.6 | 23 | Global-scale storm free | | |
| 37.8 | 29 | Global-scale storm free | | |
| 38.6 | 17 | Global-scale storm free | | |
| 44.4 | 10 | Solstice season global dust storm | 300.0 | 1.39 |
| 70.3 | 21 | Solstice season global dust storm | 254.0 | 3.09 |
| 82.4 | 28 | Solstice season global dust storm | 262.0 | 2.21 |
| 92.5 | 9 | Solstice season global dust storm | 260.0 | 2.08 |
| 92.6 | -16 | Solstice season global dust storm | 310.0 | 1.99 |
| 98.7 | 15 | Solstice season global dust storm | 208.0 | 4.46 |
| 134.5 | 27 | Global-scale storm free | | 1.83 |
| 143.7 | 1 | Solstice season global dust storm | 249.0 | 3.33 |
| 174.6 | 32 | Global-scale storm free | | |
| 213.0 | 26 | Global-scale storm free | | |
| 232.7 | 12 | Equinox season global dust storm | 204.0 | -2.05 |
| 272.1 | 31 | Global-scale storm free | | -1.24 |
| 272.4 | 25 | Equinox season global dust storm | 185.0 | -2.75 |
| 302.5 | 11 | Global-scale storm free | | -1.17 |
| 309.1 | -8 | Global-scale storm free | | -2.75 |
| 313.7 | 24 | Global-scale storm free | | -3.42 |
| 342.9 | 30 | Global-scale storm free | | |

**Table 2**. Sample of Mars years of Table 1, ordered by the phase of the *dL/dt* waveform at perihelion as determined by the method of this Section. Phase values are in degrees. *: Units of *dL/dt* amplitude are m$^6$ s$^{-2}$ M$_{Mars}$. Amplitude values are not displayed for transition years (when the *dL/dt* waveform passes through zero and changes polarity).

The *dL/dt* waveform as illustrated in Fig. 1 exhibits amplitude and phase variations that are quite unlike the regular oscillations of a sine curve. We caution that the phase value as defined here and as listed in Table 2 is a fairly crude representation of the dynamical variability with which we are concerned. While we will employ this parameter for a statistical analysis, we recognize that the phases of Table 2 should not generally be employed as a basis for high-precision modeling or dynamical calculations. The imprecision of the phase parameter is a disadvantage that must be balanced against the advantages of the phase assignment method, which include its objectivity and its simplicity, for purposes of the present exploratory investigation.

3.3. Statistical Method

With reference to the foregoing sections and Fig. 1, we immediately recognize that we require a test that is sensitive to *bimodal* distributions of phases. Our hypothesis **H1b** posits that intensification of circulatory flows should occur at two different times within the cycle, i.e., at both positive and negative extrema of the *dL/dt* waveform of Fig. 1. These times are symmetrically opposed in phase, with $\varphi_{d\mathbf{L}/dt}$ = 90° and 270°. Likewise, the opposed phase values of 0° and 180° correspond to zero-crossing times when the coupling term accelerations disappear. Fortunately, a standard test is available for assessing departures from randomness in multi-modal distributions of angular data such as that presented in column 1 of Table 2.

Schuster's Test [*Schuster*, 1897], and its close cousin, the Rayleigh Test [*Mardia*, 1972], have a long history in connection with investigations of periodicities of earthquake occurrence [*cf. Schuster*, 1897; *McMurray*, 1941; *Shlein*, 1972; *Heaton*, 1975; *Klein*, 1976; *Shirley*, 1988, and references therein; *Cochran et al.,* 2004]. In this test the angular phases $\theta_i$ of the *n* events are represented as unit vectors. The unit vectors are then summed to obtain the resultant magnitude *R* and phase φ; these are given by $R = ((A^2) + (B^2))^{1/2}$ and $\varphi = \tan^{-1}(B/A)$ respectively, where $A = \Sigma \cos \theta_i$ and $B = \Sigma \sin \theta_i$. If the angles are randomly distributed, the probability of obtaining a resultant of length *R* or greater is approximately $P_r = \exp(-R^2/n)$. Acceptable results may be obtained with remarkably small sample sizes; the Rayleigh test may be employed with samples as small as *n*=5 [*Mardia*, 1972]. In order to adapt the Rayleigh/Schuster Test to bimodal cases it is merely necessary to double the angles before summing the sines and cosines. This version is often employed for evaluating bimodal distributions associated with fortnightly tidal cycles [*cf. Klein*, 1976; *Shirley*, 1988].

Prior to employing this test it is important to ensure that the underlying population from which our samples are selected is unbiased. That is, there should be a roughly equal probability of obtaining phase values from any part of the phase range by random selection. We have verified this numerically, using methods similar to those detailed in [S15]. As an additional check, we subject the 21 angular values of Table 2 to an evaluation by this test. When this is done, we obtain a vector resultant magnitude $R$ of 4.17 (a relatively small "random walk" departure from the origin). The standard version of the test yields a random probability $P_r = 0.436$, while the bimodal version returns $P_{r2} = 0.714$. Neither of these results approach the 5% level (i.e., $P_r \leq 0.05$) of statistical significance. We repeated this exercise with a somewhat larger sample of 57 values of $\varphi_{dL/dt}$, representing the phase values at perihelion for all Mars years during the interval from 1920-2030 (see the Supplementary Data). Similar results were obtained, with $R = 9.45$, $P_r = 0.208$, and $P_{r2} = 0.566$. All of the above tests indicate that the underlying sample is largely free from systematic bias. Thus we may, with confidence, employ the bimodal version of the Schuster/Rayleigh Test to determine the statistical significance, if any, of samples of phase values that may emerge from our investigation. One caveat regarding the use of this test should be mentioned in passing. As *Heaton* [1975] has noted, it is improper to adjust or rearrange the individual observations within the sample categories following one's initial examination of test outcomes, in hopes of 'improving' the results.

## 4. Results

4.1. Relationships of $\varphi_{dL/dt}$ and global-scale dust storm occurrence and non-occurrence

The phase of the *dL/dt* waveform at perihelion for each Mars year in our sample is listed in column 1 of Table 2, as determined using the method described in Section 3.2.2. Single-year waveform phase plots for each year listed in Tables 1 and 2 are provided here in Figs. 2-5. We consider in turn the subsets of transitional years, positive polarity years, and negative polarity years.

4.1.1. Transitional years

Waveforms for 7 Mars years with phases bracketing 0° ($n=5$) and bracketing 180° ($n=2$) are illustrated in Figs. 2 and 3. No global-scale dust storms occurred in any of these transitional years. Under the present physical hypothesis, a period of 'relaxation' of driven circulatory flows (as contrasted with an intensification; [S16]) should be associated with the transitional cases. Thus the non-occurrence of GDS during these 7 Mars years is consistent with the first hypothesis under investigation (**H1a**), as the coupling term accelerations of Eq. 1 diminish and disappear under these conditions, to subsequently re-emerge with a different sign.

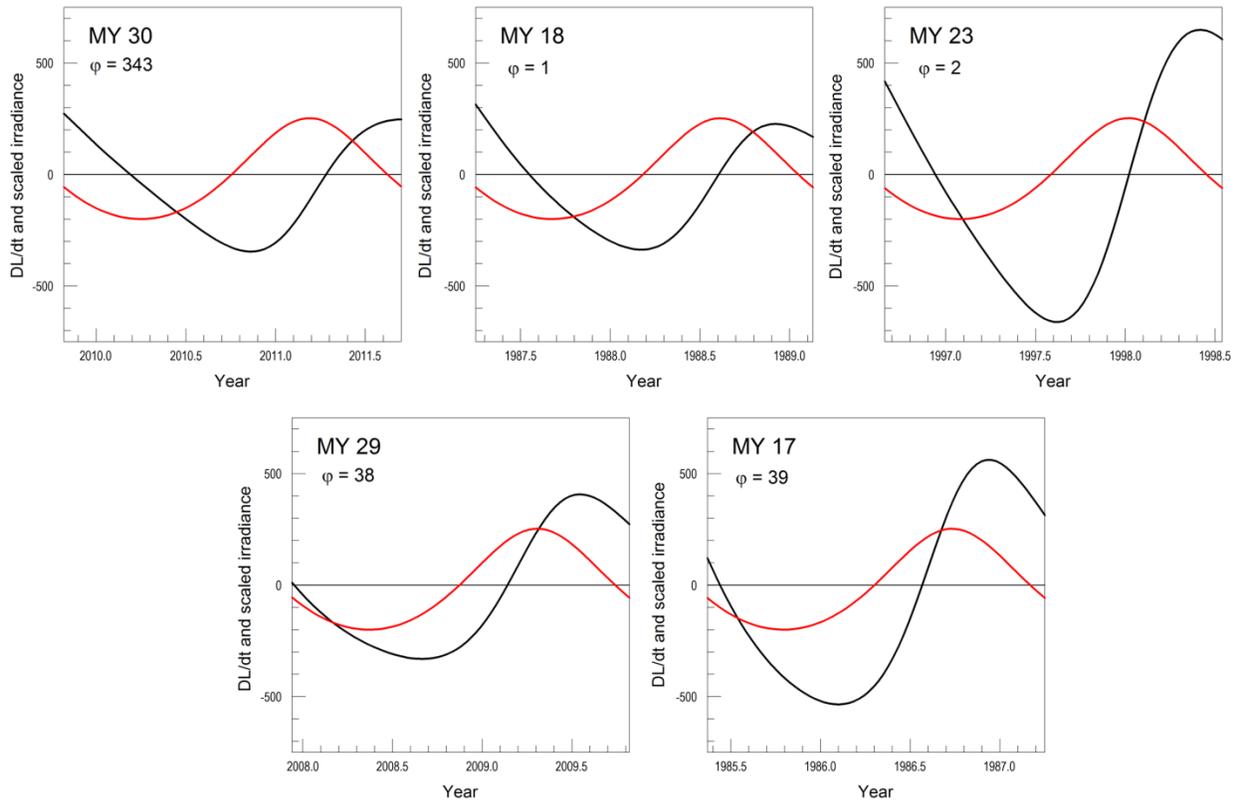

**Figure 2**. Transitional ("zero-crossing") Mars years with phases bracketing 0°. (A phase value of precisely 0° would be obtained if the transition from negative to positive values of *dL/dt* occurred concurrently with Mars reaching perihelion). Each panel extends over one Mars year. The solar irradiance is indicated in red, while the *dL/dt* waveform is represented in black. Maximum values of the solar irradiance are attained at perihelion, which occurs in the second half of the Mars year. Both curves have been scaled to better illustrate the relationships. No global-scale dust storms occurred in any of these years.

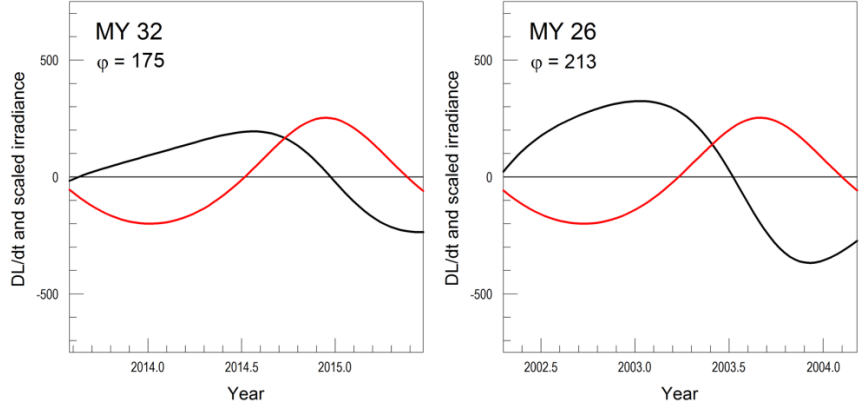

**Figure 3**. Transitional ("zero-crossing") Mars years with phases bracketing 180°. Each panel extends over one Mars year. The solar irradiance is indicated in red, while the *dL/dt* waveform is represented in black. Maximum values of the solar irradiance are attained at perihelion, which occurs in the second half of the Mars year. Both curves have been scaled to better illustrate the relationships. No global-scale dust storm occurred in either of these years.

4.1.2. Positive polarity Mars years

Shown in Fig. 4 are the 8 Mars years of Table 2 with $\varphi_{dL/dt}$ ranging from 44°-144°. A common feature of the waveform phase plots of Fig. 4 is the occurrence of positive extrema of the *dL/dt* waveform in the second half of the Mars year. In keeping with the prior discussion of [S16] and Section 3.2, we will refer to these examples as "positive polarity" years.

Global-scale dust storms occurred in 7 of the 8 Mars years plotted in Fig. 4, with the sole exception being Mars year 27 (with perihelion occurring in 2005). *All of the southern summer solstice storms of Table 2 are included here.* We note that all of the *dL/dt* waveforms of Fig. 4, with the exception of MY 27, are concave-upward in the first half of the year. The waveform for MY 27 is likewise distinguished from the others by its long rise time and moderate slope. Further discussion of the exceptional characteristics of MY 27 is deferred to Section 5.1.

Characteristics of the outlying Mars years with the lowest ($\varphi_{dL/dt}$ =44°) and highest ($\varphi_{dL/dt}$ =144°) positive polarity phase values are also discussed in greater detail in Section 5.1 below.

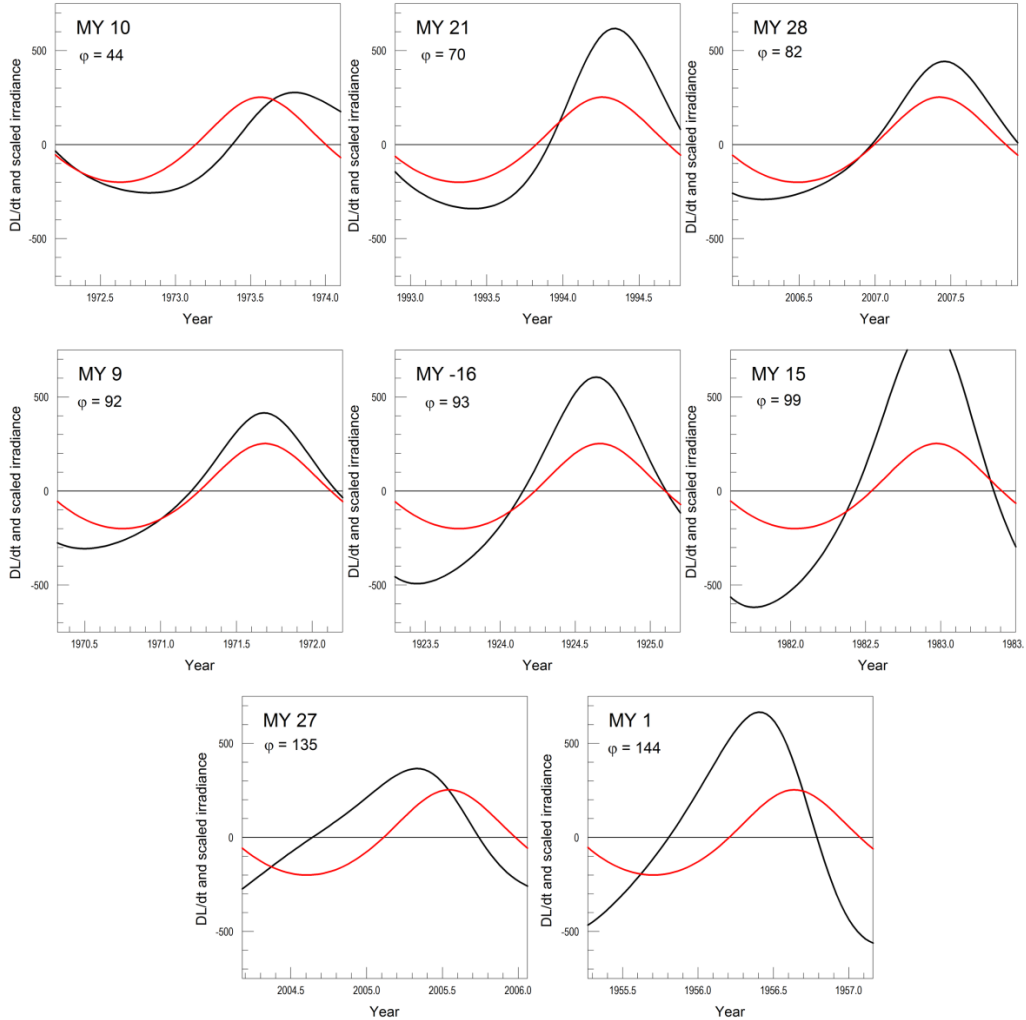

**Figure 4**. Positive polarity Mars years. A positive peak of the *dL/dt* waveform is seen during the second half of the Mars year in these cases. Phase values increase from left to right and from top to bottom. Each panel extends over one Mars year, with the solar irradiance indicated in red, while the *dL/dt* waveform is shown in black. Global-scale dust storms occurred in 7 of the 8 Mars years illustrated, with the exception of MY 27, at lower right. All of the southern-summer-

solstice season global-scale dust storm years of Table 1 are included here. Both curves have been scaled to better illustrate the relationships.

With respect to the linked physical hypotheses under evaluation here, we recall from Section 3.2 that extrema of the *dL/dt* waveform correspond to the midpoints of the predicted intensification intervals for the global circulation. The occurrence of all 7 of the southern summer solstice season global-scale dust storms of Table 1 during positive polarity years is consistent with **H1b**.

4.1.3. Negative polarity Mars years

The 6 Mars years of Table 2 with $\varphi_{dL/dt}$ ranging from 233°-313° are illustrated (in ascending order of phase) in Fig. 5. Note that the transition from positive to negative values of the *dL/dt* waveform occurs progressively earlier in the Mars year in each plot. In all cases, negative extrema of the *dL/dt* waveform were attained in the second half of the Mars year; thus these are identified as negative polarity Mars years. Substantial variability of the waveform amplitude is observed here, as in Figs. 2 and 4.

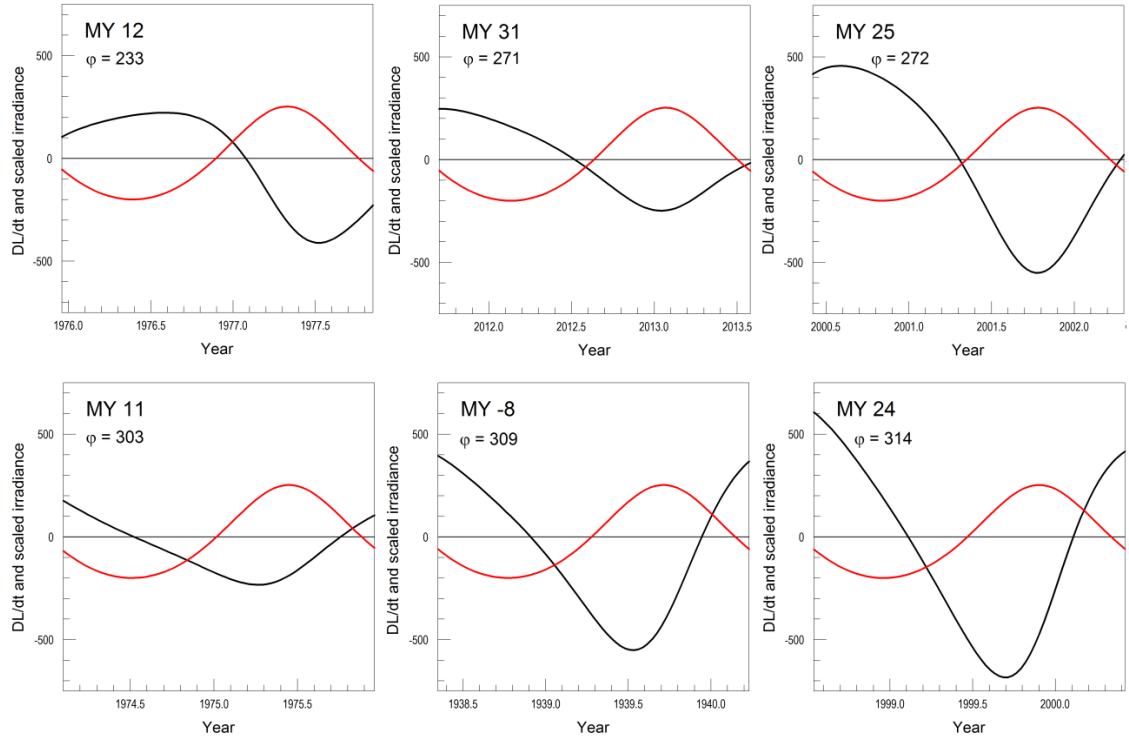

**Figure 5**. Negative polarity Mars years, with $\varphi_{dL/dt}$ ranging from 233°-314°. Phase values increase from left to right and from top to bottom. Each panel extends over one Mars year, with the solar irradiance indicated in red, while the *dL/dt* waveform is indicated in black. Equinox season global dust storms occurred in MY 12 and MY 25, while the remaining 4 Mars years illustrated were global-storm-free.

Under **H1b**, the condition with $\varphi_{dL/dt}$ =270°, like that with $\varphi_{dL/dt}$ =90°, is predicted to give rise to an intensification of circulatory flows. Of the 6 Mars years shown with phases between 233° and 313°, two (MY 12 and MY 25) were accompanied by GDS. Each of these storms occurred exceptionally early in the dust storm season; they are accordingly labeled 'equinox season global dust storms' in Table 1. Prior to the publication of [S15], no common factors had previously been found linking these two anomalous years one with the other. As in [S15], we note a fair correspondence in phase of the putative forcing function linking the anomalous GDS years MY 12 ($\varphi_{dL/dt}$ =233°) and MY 25 ($\varphi_{dL/dt}$ =272°). Further discussion regarding the four global-storm-free negative polarity Mars years is deferred to Sections 4.2 and 5.2 below. We next address the statistical significance of the relationships uncovered.

4.2. Statistical evaluation

In Section 3.3 above we employed the bimodal version of Schuster's Test to evaluate the sample of $\varphi_{dL/dt}$ values for all of the Mars years of Table 2, finding no evidence of nonrandom clustering of phases for our complete sample. We now subdivide the catalog of 21 Mars years of Table 2 into two samples for evaluation: The global dust storm years, with $n$=9, and the years lacking such storms, with $n$=12. Phase values comprising each sample are illustrated in Fig. 6.

Considering first the distribution for the global-scale dust storm years (Fig. 6a), we immediately note that the phases for the equinox season GDS of Mars years 12 and 25 (in blue) are approximately diametrically opposed to the phases obtained for the 7 southern summer solstice GDS years (in orange). The phases for the complete sample of GDS years are seen to cluster about an axis between phases ~90° and ~270°, corresponding to the phases of the positive and negative extrema of the $dL/dt$ waveform of Fig. 1. Application of Schuster's Test yields a random probability of occurrence for the global-scale dust storm years of $P_{r2}$ = 0.024, thereby attaining statistical significance at better than the 5% level. Distributions with a similar degree of bimodal clustering would emerge through random sampling of the underlying distribution about once in every 40 trials.

This statistical result confirms hypothesis **H1b**, as stated in Section 3.1: We have found a direct relationship linking global-scale dust storm occurrence on Mars with the timing of the predicted circulatory intensification cycles linked with the forcing function *dL/dt*.

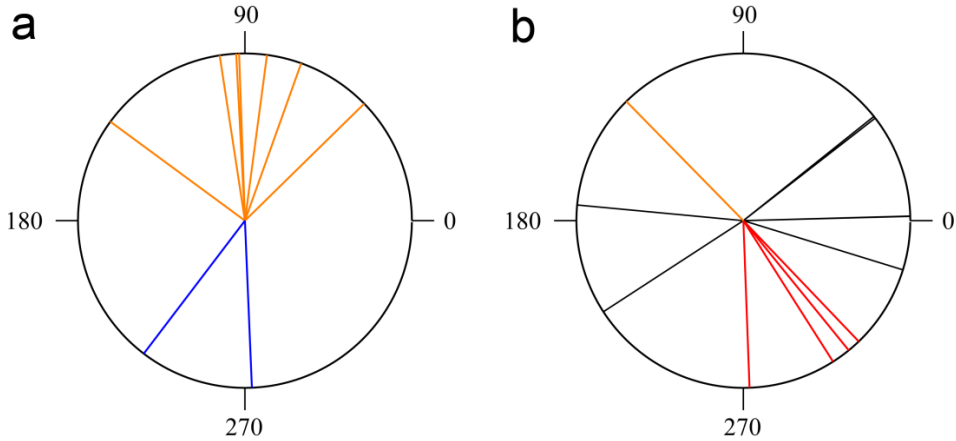

**Figure 6**. Phases ($\varphi_{dL/dt}$) for Mars years with global-scale dust storms (left) and for Mars years without global-scale dust storms (right). **a**): *dL/dt* phases for southern summer solstice season GDS years (Table 2) are indicated in orange, while phases for equinox season GDS years are indicated in blue. The (bimodal) random probability of occurrence for this distribution is $P_{r2}=0.024$. **b**): $\varphi_{dL/dt}$ for Mars years lacking GDS. The distribution does not attain statistical significance, with $P_{r2}=0.367$. Three populations are identified: In black are the 7 'transitional' Mars years of Figures 2 and 3, with $\varphi_{dL/dt}$ in all cases lying within 40° of the zero-crossing phases (0° and 180°). Nearly coincident phases values were obtained for MY 18 (1.5°) and MY 23 (1.6°), and for MY 17 (38.6°) and MY 29 (37.8°). Only a single radial vector is discernible in the plot for each of these nearly overlapping cases. In red, with phases between 271° and 314°, are 4 GDS-free negative polarity Mars years (Fig. 5). The phase for Mars year 27, the only storm-free year of Table 2 with positive polarity, is shown in orange.

The distribution of phases for the 12 Mars years lacking global-scale dust storms (Fig. 6b) does not exhibit notable clustering and returns a random probability estimate of $P_{r2}$=0.367, which is not statistically significant.

The plotted phases of Fig. 6b appear to represent two separate bimodal distributions, with differing preferred directions. Shown in black in Fig. 6b are the 7 'transitional' years of our sample, whose plotted phases identify a preferred axis corresponding approximately to the 0° and 180° directions. Also shown in Fig. 6b (in red) are phases for the 4 negative polarity storm-free years of Fig. 5, together with the single positive polarity storm-free year MY 27, shown in orange, which is opposed in phase with respect to the negative polarity set. The (bimodal) preferred direction (or "axis") in this case appears to be offset by 40° or so with respect to the transitional years distribution.

Because we have no *a priori* basis for distinguishing between these subsets of the catalog of Mars years without GDS (i.e., the transitional years, versus the storm-free positive and negative polarity years), we cannot fairly test for or claim statistical significance for either subset. Even so, the pattern of the GDS-free positive and negative polarity years may be of some interest. The phases for all 5 of these Mars years correspond to periods that are past the putative peak acceleration epochs (which occur at phases of 90° and 270°). The calculated coupling term acceleration magnitudes are therefore decreasing, at perihelion, in each of these years. Our sample is small, but the patterns of Fig. 6 nonetheless suggests that intervals when the phase is approaching and near 90° and 270° may be somewhat more favorable for the inception of GDS than is the case for positive and negative polarity years with phase values significantly greater than 90° and 270° respectively.

## 5. Discussion

5.1 Outliers, anomalies, and the role of the waveform amplitude

In prior sections we have been concerned with uncovering common factors that link the years with GDS (and the years without GDS) one with another. To this point we have deliberately avoided discussing additional characteristics (other than the phase of the *dL/dt* waveform) with respect to individual years. Consequently the waveform amplitude, as tabulated in Table 2, has been mentioned only in passing. We are now in a position to look more systematically at the outliers and possibly discrepant cases uncovered thus far.

Substantial variability of the waveform amplitude is evident in Fig. 1 and in Figs. 2-5. It is natural to suspect that the waveform amplitude should play a role in determining whether or not a GDS may occur in any given year. A consideration of waveform amplitudes of $L_{Mars}$ in [S15], however, was unable to reach any definite conclusions (*cf.* Fig. 13 of [S15]). While a number of GDS years exhibited substantial $L_{Mars}$ waveform amplitudes, others (notably MY 10 and MY 12) did not. Our investigation differs from the prior study, in that *dL/dt* has here been directly linked with a specific physical forcing mechanism ([S16]; Section 3.1). It thus makes sense to re-examine the question of waveform amplitudes here.

We first consider the case of the negative polarity Mars year 31, in juxtaposition with MY 25. The phase for MY 31, at 271°, is nearly identical to that found for MY 25 (272°), which is a GDS year. Why was no comparable storm observed in MY 31? In Table 2 we note that the amplitude of the *dL/dt* waveform is much reduced (by more than a factor of 2) for MY 31 in comparison to that for MY 25. The reduced magnitude of the resulting coupling term accelerations may conceivably help to account for the non-occurrence of a global-scale dust storm in this otherwise favorably phased year; among other possibilities, it may be that some critical threshold value, perhaps for surface wind stress, was not attained.

Reduced waveform amplitude may also provide a possible explanation for the non-occurrence of a global-scale dust storm in MY 27, a positive polarity year with $\varphi_{dL/dt}=135°$. This is the only positive polarity Mars year of Fig. 4 lacking such a storm. In Table 2, we note that MY 27 has the second-lowest waveform amplitude of any positive polarity year. Enthusiasm for this explanation must be tempered, however, by the realization that the amplitude for the 1973 GDS year (MY 10), discussed next, was smaller still.

The paired global-scale dust storms of MY 9 and 10 have puzzled Mars atmospheric scientists for a considerable time. This is the only known example of GDS occurring in two successive Mars years. This recurrence may (or may not) present difficulties for conceptual models in which control over GDS occurrence is determined by the availability (or unavailability) of surface dust reservoirs in key locations (Section 1; see also *Newman and Richardson* [2015]). However, the occurrence of GDS in successive Mars years is not a difficulty for the present hypothesis. As indicated in Table 2 and Fig. 4, positive polarity phasing of the *dL/dt* waveform (favorable for the occurrence of GDS) is present in both years.

MY 10 is unusual in another way as well. As remarked earlier, the phase value (with $\varphi_{dL/dt}=44°$) is an outlier, being the lowest for any of the positive polarity GDS years. Inspection of the MY 10 panel of Fig. 4 indicates that (as is appropriate for this phase), the peak value of the forcing function is not attained until very late in the year. As indicated in Table 1, the 1973 GDS was initiated quite late in the dust storm season, at $L_s=300°$. While the delayed onset in MY 10 seems unlikely to be coincidental, we cannot draw strong conclusions on the basis of a sample of one. MY 10 may thus represent an attractive and challenging subject for numerical modeling.

The phase of the global-scale dust storm of MY 1 (1956), with $\varphi_{dL/dt}=144°$, likewise represents an outlier with respect to the other positive polarity Mars years of Fig. 4. This phase lies closer to the nearest zero-crossing phase (180°) than to the positive peak phase (90°). In this case, the peak value of the *dL/dt* waveform was attained very early in the dust storm season, while the storm is believed to have initiated very close to the time of perihelion, at $L_s=249°$. We note in Table 2 that the waveform amplitude for MY 1 is the second largest of any of the positive polarity years. With this phase value, under the orbit-spin coupling hypothesis, a marked intensification of circulatory flows would be expected to occur in the southern spring, extending thereafter well into the first half of the dust storm season. We suspect that the large waveform amplitude is what sets this year apart from MY 27, which exhibits a neighboring phase value (with $\varphi_{dL/dt}=135°$), but in which no global-scale dust storm occurred.

While Mars year 15, with perihelion in 1982, has not previously been singled out for attention, it is nonetheless noteworthy for showing by far the largest waveform amplitude of any of the positive polarity Mars years of our sample. The phase (99°) indicates that the peak acceleration would have occurred a short time following perihelion. However, the GDS of 1982 in fact began much earlier in the dust storm season, near $L_s=208°$ (Table 1; [S15]). Here we may likewise appeal to the large waveform amplitude as a possible explanation for the early inception date of the global storm. As indicated in Fig. 4, the waveform amplitude for this case is so large as to exceed the peak values attained in most of the other positive polarity years quite early in the 1982 dust storm season.

The preceding discussion serves to once more emphasize the limitations of the phase parameter, considered in isolation, as a metric for characterizing the dynamical variability of the *dL/dt* waveform as shown in Figs. 1-5 of this paper. By taking account of the waveform amplitude, together with the phase, we have been able to provide some plausible but speculative scenarios for the observed outcomes for MY 1, 10, 15, 27, and 31. Global circulation modeling provides more satisfactory explanations for the occurrence and non-occurrence of GDS in some of these Mars years [*Mischna and Shirley*, 2016; this issue].

5.2. Polarity and global-scale dust storm occurrence

Seven of the eight positive polarity Mars years of Table 2 were accompanied by global-scale dust storms. The relationship uncovered is consistent with hypothesis **H1b**. The situation is less straightforward, however, in the case of the negative polarity years (Fig. 5). While our physical hypothesis predicts an intensification of the global circulation for these years, as in the case of the positive polarity years, only two of the six negative polarity Mars years were accompanied by global-scale dust storms. While the occurrence of two equinox-season storms during negative polarity intensification intervals is consistent with **H1b**, the presence of four other storm-free Mars years with negative polarities suggests that our initial working hypotheses of Section 3.1.3 may be overly simplified or may be otherwise lacking in sophistication.

We emphasize that there is one quite fundamental difference between the positive and negative polarity cases that is not adequately addressed under **H1b**. In positive polarity intervals, Mars is gaining orbital angular momentum at the expense of other members of the solar system family; while in negative polarity intervals, the opposite condition is true [S15, S16]. In consequence, all else being equal, *the directions of the coupling term accelerations are everywhere reversed* during negative polarity intervals, in comparison with those of positive polarity cases [S16]. Thus it is not unreasonable to suppose that the resulting motions of the atmosphere may thus take on quite different forms during intensification intervals of differing polarities. Numerical modeling [*Mischna and Shirley*, 2016] has begun to illuminate the differences in the effects of the coupling term accelerations during negative polarity episodes. Further exploration of this important question may be found in a companion paper [*Mischna and Shirley*, 2016].

## 6. Caveats and some physical implications

While the present investigation has uncovered circumstantial evidence that supports the physical reality of the orbit-spin coupling mechanism of [S16], it is important to recognize that the preliminary investigation reported here does not constitute a critical experiment (i.e., one that is capable of unambiguously confirming or disqualifying the physical hypothesis).

A second important caveat is that we have implicitly assumed that an overall intensification of the circulation of the Mars atmosphere can lead directly to the occurrence of global-scale dust storms. While plausible, this linkage is as yet unproven. (New evidence bearing on this question is described in *Mischna and Shirley* [2016]). GDS on Mars may result from some other mechanism (or combination of mechanisms) entirely. Further, a portion of the evidence offered is statistical in nature, even though the questions asked and the comparisons performed conform to and are dictated by a specific physical model.

The occurrence of global-scale dust storms on Mars in some years but not in others has been characterized as an important unsolved problem of the physics of planetary atmospheres [*Haberle*, 1986, *Kahn et al.*, 1992, *Zurek et al.*, 1992, *Zurek and Martin*, 1993, *Pankine and Ingersoll*, 2002, 2004; *Basu et al.*, 2006; *Cantor*, 2007. Our results suggest that the correspondence between model outcomes and reality may be improved by including orbit-spin coupling accelerations within our numerical models. A first step in this direction is reported in the companion paper by *Mischna and Shirley* [2016].

## 7. Forecasts for the dust storm seasons of Mars Years 33 and 34

The dust storm season of the current Mars year (MY 33), as defined in [S15], will begin on 1 June 2016 ($L_s=160°$) and will end on 28 March 2017 ($L_s=340°$). Mars will reach perihelion in late October 2016. As indicated in Fig. 1, the amplitude of the *dL/dt* waveform during the dust storm season of MY 33 will be approximately twice as large as at any time during the preceding 3 Mars years. The positive peak of the *dL/dt* waveform occurs a short time before Mars' perihelion; we accordingly obtain a phase value of 105° for MY 33, using the method of Section 3.2. This phase falls within the positive polarity phase range displayed in Figs. 4 and 6a, which includes the phases for all seven of the known southern summer solstice season global-scale dust storms of Table 2. On the basis of the correspondence shown in Fig. 7 we believe that a global-scale dust storm is likely to occur during the southern summer dust storm season of Mars year 33.

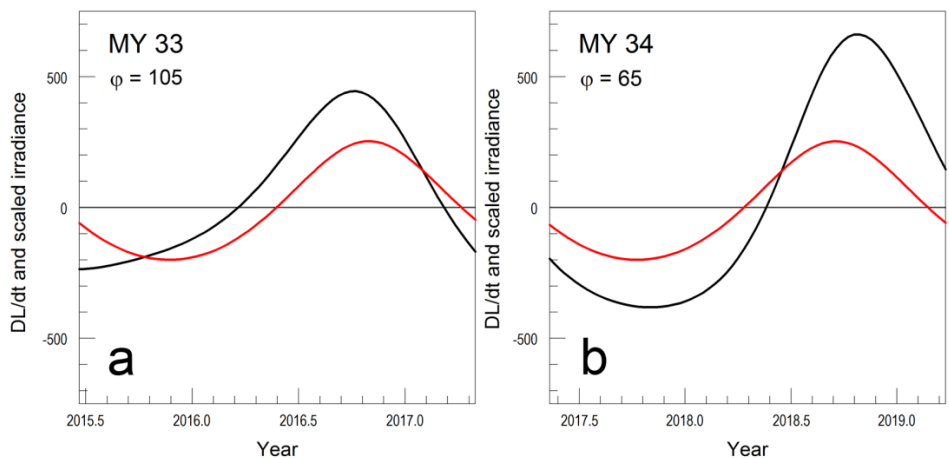

**Figure 7**. **a**). Positive polarity *dL/dt* waveform for the current Mars year (MY 33, with $\varphi_{dL/dt}$ =105°), and **b**) the waveform for MY 34 (with $\varphi_{dL/dt}$ =65°). These may usefully be compared with the positive polarity waveforms of Fig. 4.

In Fig. 7 we compare the waveform for MY 33 with that for MY 34 ($\varphi_{dL/dt}$ =65°). The waveform amplitude for MY 34 is large in comparison with that for MY 33, and the phase, at 65°, makes this also a member of the positive polarity class. Based on the results for positive polarity Mars years described above, we conclude that conditions favorable for the occurrence of a GDS are also present for the dust storm season of 2018 / MY 34. While the occurrence of GDS in these two successive years is a possibility, as previously observed in MY 9 and MY 10, we cannot be entirely confident that this will take place. The redistribution of dust in a MY 33 storm might prevent the occurrence of a global-scale storm in MY 34 [*Mulholland et al.*, 2013; *Newman and Richardson*, 2015]. If no storm occurs in MY 33, then we would have greater confidence in forecasting a GDS for MY 34, based on the significantly larger waveform amplitude of MY 34.

A similar pair of forecasts was advanced in [S15], following a comparison of waveforms of the orbital angular momentum of Mars for different Mars years. While the forecasts are in many respects the same, we believe that the present forecast rests on a stronger foundation, as a testable physical hypothesis now informs the analysis.

**8. Conclusions**

The orbit-spin coupling hypothesis of *Shirley* [2016] predicts that cycles of intensification and relaxation of circulatory flows of atmospheres will occur in phase with the variability of the solar system barycentric orbital angular momentum $d\mathbf{L}/dt$. This hypothesis is confirmed (albeit indirectly), for the case of the Mars atmosphere, through a comparison of the phasing of the $d\mathbf{L}/dt$ waveform (with respect to the southern summer dust storm season on Mars) with the historic record of occurrence and non-occurrence of global-scale dust storms on Mars. No global-scale dust storms occurred on Mars during seasons associated with the diminution and disappearance of the putative forcing function $d\mathbf{L}/dt$. All of the known global-scale dust storms occurred during seasons in which the physical hypothesis predicts intensification.

This investigation marks the first-ever comparison of the forcing function $d\mathbf{L}/dt$ with atmospheric variability for any planet. This investigation likewise represents the first direct and explicit test of the orbit-spin coupling hypothesis of *Shirley* [2016]. While our investigation does not constitute a critical experiment, the present results nonetheless provide strong support for the physical reality of the specific coupling mechanism proposed. While we cannot conclude that the phenomenon of global-scale dust storm occurrence on Mars is controlled solely by this dynamical mechanism, the results do strongly indicate that orbit-spin coupling is likely to play an important role in the excitation of interannual variability of the Mars atmosphere. This conclusion has wider implications, particularly for the presently open question of the origins of seasonal and longer-term weather and climate anomalies and variability on Earth. A companion paper [*Mischna and Shirley*, 2016] describes initial results of numerical modeling investigations of the interannual variability of the Mars atmosphere with orbit-spin coupling accelerations.

**10. Acknowledgements**

The authors acknowledge the support of NASA's Solar System Workings Program (14-SSW14_2-0354) and JPL's Research and Technology Development Program. The authors have benefited from critical feedback and discussions with many individuals inside and outside JPL, including Rich Zurek, Don Banfield, Bruce Cantor, Bruce Bills, Tim Schofield, David Kass, Armin Kleinböhl, Nick Heavens, Dan McCleese, Paul Hayne, and other members of the MRO-MCS science and operations teams. This work was performed at the Jet Propulsion Laboratory, California Institute of Technology, under a contract from NASA. Datasets employed for this investigation are provided in the online Supporting Information.